\documentclass[12pt,preprint]{aastex}

\slugcomment{Accepted by ApJ at August 21, 2014.}

\shorttitle{Induced Core Formation Time in Subcritical Magnetic Clouds}
\shortauthors{Kudoh \& Basu}

\begin{document}


\title{Induced Core Formation Time in Subcritical Magnetic Clouds 
    by Large-Scale Trans-Alfv\'enic Flows}


\author{Takahiro Kudoh\altaffilmark{1}}
\affil{Division of Theoretical Astronomy, National Astronomical Observatory of Japan,
2-21-1 Osawa, Mitaka, Tokyo 181-8588, Japan}
\email{kudoh@th.nao.ac.jp}

\and

\author{Shantanu Basu}
\affil{Department of Physics and Astronomy, University of Western Ontario, London,
Ontario N6A 3K7, Canada}
\email{basu@uwo.ca}


\altaffiltext{1}{Department of Astronomical Science, School of Physical Sciences, 
  the Graduate University for Advanced Studies (SOKENDAI), 
  2-21-1 Osawa, Mitaka, Tokyo 181-8588, Japan}


\begin{abstract}
We clarify the mechanism of accelerated core formation by large-scale
nonlinear flows in subcritical magnetic clouds by finding
a semi-analytical formula for the core formation time and
describing the physical processes that lead to them. 
Recent numerical
simulations show that nonlinear flows induce
rapid ambipolar diffusion that leads to localized supercritical regions
that can collapse. Here, we employ
non-ideal magnetohydrodynamic simulations including ambipolar diffusion
for gravitationally stratified sheets threaded by vertical magnetic fields.
One of the horizontal dimensions is eliminated, resulting in a simpler
two-dimensional simulation that can clarify the basic process of 
accelerated core formation. 
A parameter study of simulations shows
that the core formation
time is inversely proportional to the square of the flow speed when the
flow speed is greater than the Alfv\'en speed. 
We find a semi-analytical formula that explains this numerical result.
The formula also predicts that the core formation time is about three times 
shorter than that with no turbulence, when the turbulent speed is comparable 
to the Alfv\'en speed. 
\end{abstract}





\section{Introduction}

The theory that the star formation process in a molecular cloud is controlled by 
the magnetic field and ambipolar diffusion has been studied for many years
\citep[e.g.,][]{mou78,shu87}.
In the theory, the mass-to-flux ratio, which corresponds to the relative strength 
of gravity and the magnetic field, is an important parameter for the process.
If the mass-to-flux ratio is greater than its critical value 
(i.e., supercritical : gravity dominates the magnetic field),
the cloud is likely to collapse or fragment on the dynamical timescale.
If the mass-to-flux ratio is less than its critical value 
(i.e., subcritical : the magnetic field dominates gravity), 
the gravitational driven collapse or fragmentation occurs on the ambipolar 
diffusion timescale \citep[e.g.,][]{lan78,zwe98} instead of the dynamical timescale
as long as the diffusion timescale is longer than the dynamical timescale.
Although there are some observational difficulties to determine whether 
the mass-to-flux ratio of molecular clouds are greater or lesser than 
the critical value prior to core formation 
\citep[e.g.][]{cru09,mou09},
subcritical molecular clouds are consistent with
some observations that show that molecular clouds are associated with large-scale 
ordered magnetic fields 
\citep[e.g.,][]{cor05,hey08,alv08,li09}.

Here, we focus on fragmentation and core formation in subcritical molecular clouds.
The first numerical simulations of fragmentation in subcritical clouds 
with ambipolar diffusion were performed by \citet{ind00} under 
the assumption of a two-dimensional infinitesimally thin sheet. 
\citet{bas04} carried out simulations
of magnetized sheets including the effect of a finite disk half-thickness 
that was consistent with hydrostatic equilibrium. 
A three-dimensional simulation of fragmentation
and core formation in subcritical clouds with ambipolar diffusion, 
including full gravitational stratification along the magnetic field,
was first performed by \citet{kud07}.
These numerical calculations confirmed that the gravitationally driven fragmentation 
and core formation in subcritical clouds occur on the ambipolar diffusion timescale. 
Core formation in subcritical clouds generally explains 
the low star formation rate in our galaxy,
since the typical ambipolar diffusion time is more than 10 times 
longer than the dynamical time in molecular clouds.
On the other hand, a problem has been pointed out 
that the ratio of starless and stellar cores is smaller
than that is expected from the theory \citep[e.g.,][]{jij99}.
The observations mean that core formation is likely to occur on up 
to several dynamical times, i.e., the usual ambipolar diffusion time is 
about $3-10$ times longer than that expected from the observations.

Ideas to solve the timescale problem were proposed by \citet{fat02}, 
who pointed out that fluctuating magnetic fields and density in turbulent 
molecular clouds can enhance the ambipolar diffusion rate
because the ambipolar diffusion is a process of nonlinear diffusion.
Their analysis indicated that the density fluctuation enhances 
the diffusion rate more efficiently when the amplitude of 
the fluctuation is large.
Their idea is consistent with the fact that supersonic turbulence
is considered to be a significant factor in molecular clouds
as well as for magnetic fields during the early stage of star formation
\citep[e.g.,][]{mck07}, since the supersonic turbulence tends to enhance the
density through compression.
\citet{li04} carried out numerical simulations of
fragmentation in subcritical clouds with a thin-sheet approximation, 
including both ambipolar diffusion and 
supersonic turbulent flow. They found that the timescale of core formation 
in subcritical clouds is reduced by the supersonic turbulent flow as long as 
the turbulence is dominated by large-scale fluctuations. 
\citet{bas09} confirmed the fast core formation by nonlinear large-scale
turbulent flows with a numerical simulation using the thin-sheet approximation. 
They also found that the subcritical clouds experienced rebounds after 
the first compression and showed several oscillations before 
the core formation as long as the initial flow speed is trans-Alfv\'enic.
These results were also confirmed by 
\citet{kud08} and \citet{kud11} 
through three-dimensional 
simulations without the thin-sheet approximation.
\citet{kud11} found that the core formation time ($t_{core}$) roughly scales as
$t_{core} \propto \rho_{peak}^{-0.5}$, where $\rho_{peak}$ represents
the first density peak created by the compression induced by the
supersonic turbulent flow. 
\citet{kud11} also confirmed that the core formation time is
shorter when the turbulent flow speed is greater. 
Although these results indicate that the density enhancement in 
subcritical clouds is an important factor to shorten 
the core formation time,
they did not clarify the physical process of how the density enhancement shortens 
the core formation time as well as how the core formation time
depends on the turbulent flow speed.
This motivates us to carry out further investigations of the induced core formation 
by nonlinear flows in subcritical clouds.
Previous studies indicated that large-scale compression would be 
the key process to enhance the ambipolar diffusion rate. 
In this short paper, we present the investigation of the core formation 
under a simple large-scale nonlinear flow rather than complex turbulent flows.
We study basic properties of the core formation in subcritical clouds 
to achieve a clear explanation of how the fast ambipolar diffusion 
is induced by large-scale compression in the clouds.

This paper is  organized as follows.
We have a brief description of the setup for numerical simulations in Section 2. 
The numerical results and its semi-analytical explanation are given in Section 3. 
We summarize the results and have some discussions in Section 4.

\section{Setup for Numerical Simulation}

The numerical setup in this paper is similar to that used in 
\citet{kud11}, 
but spatial symmetry is assumed in one of the directions
perpendicular to the magnetic field.
We effectively solve the magnetohydrodynamic (MHD) 
equations in two spacial dimensions with self-gravity and ambipolar diffusion.
The equation for the time evolution of the magnetic field 
with ambipolar diffusion is assumed to be the one-fluid model as
\begin{equation}
\frac{\partial \mbox{\boldmath$B$}}{\partial t}
= \nabla \times (\mbox{\boldmath$v$} \times \mbox{\boldmath$B$})
+ \nabla \times \left[\frac{\gamma}{4\pi \rho^{3/2}} (\nabla \times \mbox{\boldmath$B$}) \times \mbox{\boldmath$B$}) 
\times \mbox{\boldmath$B$}\right], 
\label{eq:ind}
\end{equation}
where {\boldmath$B$} is the magnetic field,
{\boldmath$v$} is the velocity, 
$\rho$ is the density of neutral gas, 
$\gamma \simeq 170.2$ g$^{1/2}$cm$^{-3/2}$ s is used as the typical value
in molecular clouds
\citep[e.g.,][]{bas94}.
In the two-dimensional Cartesian coordinate system $(x,z)$, 
the initial magnetic field is assumed to be uniform along the $z$-direction:
\begin{equation}
B_z=B_0,\ \ B_x=0,
\end{equation}
where $B_x$ and $B_z$ are the magnetic field components of the $x$- 
and $z$-directions, respectively, and $B_0$ is constant.
In this coordinate system, we assume physical variables are constant
along the direction perpendicular to both the $x$- and $z$-directions.

The initial hydrostatic equilibrium of a self-gravitating 
cloud along the magnetic field ($z$-direction) is assumed.
If the temperature of the cloud is uniform throughout the region,
the analytic density distribution $\rho_S$ was obtained by \citet{spi42}:
\begin{equation}
\rho_S(z)=\rho_0 \,\mbox{sech}^2(z/H_0),
\label{eq:sp}
\end{equation}
where
\begin{equation}
H_0=\frac{c_{s0}}{\sqrt{2\pi G \rho_0}},
\label{eq:h0}
\end{equation}
$\rho_0$ is the density at $z=0$, 
$c_{s0}$ is the sound speed, and $G$ is the gravitational constant.
Instead of the uniform temperature, we use the hyperbolic tangent 
function for temperature distribution that shows a sharp transition
from low temperature to 10 times higher temperature around $z=2H_0$
\citep[see][]{kud11}.
This function mimics the situation that warm gas commonly surrounds
the isothermal molecular cloud.
The initial hydrostatic density distribution obtained from the temperature distribution 
is almost the same as that of equation (\ref{eq:sp}) in the low temperature region
in which we are interested.
We model molecular clouds in the number density 
range $10^3$ cm$^{-3} - 10^6$ cm$^{-3}$. 
Since the the cooling time is much shorter than the dynamical time
in this density range, we adopt isothermality
for each Lagrangian fluid element \citep{kud03,kud06}.

In the sheet we 
mimic the largest mode of a turbulent flow with an initial sinusoidal 
velocity fluctuation along the $x$-direction: 
\begin{equation}
	v_x= - v_{max} \sin (2\pi x /x_{max}) \ \ \ \ \ \mbox{(if $|x| \leq 0.5x_{max})$}
	,
	\label{eq:iniv}
\end{equation}
where $v_x$ is the velocity component of $x$-direction, $v_{max}$ is the 
velocity amplitude of the fluctuation and a variable parameter in this work,
and $x_{max}$ is fixed to equal $4 \pi H_0$ throughout the paper. 
The most unstable wavelength of gravitational instability 
is about $4\pi H_0$ in the case with no magnetic field \citep{miy87a},
and the corresponding wavelength with magnetic field is
expected to be close to $4\pi H_0$ unless the cloud is 
marginally critical \citep{cio06}. 
Due to this sinusoidal velocity fluctuation, the initial flow 
converges the cloud toward $x=0$.
Since the symmetry is assumed in the direction perpendicular to the $x$- and $z$-directions,
the converging flow makes a filamentary structure rather than a spherical one.
In the previous three-dimensional simulations of large-scale turbulent flows 
with energy spectrum of $k^{-4}$ (Kudoh \& Basu 2011), 
where $k$ is the wave number of the turbulence,
a similar filamentary structure is produced in the cloud before 
the gravitational fragmentation.
The sinusoidal velocity fluctuation is a simple model to simulate 
the fragmentation and collapse in a large-scale turbulent flow.
 
As units for this problem, 
$H_0$, $c_{s0}$, and $\rho_0$ 
are chosen for length, velocity, and density, respectively.
These give a time unit $t_0 \equiv H_0/c_{s0}$.
The strength of the initial magnetic field
brings one dimensionless free parameter,
\begin{equation}
\beta_{0}  \equiv  \frac{8 \pi \rho_0 c_{s0}^2}{B_0^2}.
\end{equation}
The parameter $\beta_{0}$ represents the initial ratio of gas to magnetic pressure 
at $z=0$, and has a relation with $\mu_S$ as
\begin{equation}
\beta_{0}=\mu_S^2,
\label{eq:bms}
\end{equation}
where $\mu_S$ is the the mass-to-flux ratio normalized to the critical value
of the cloud whose density distribution is described by equation 
(\ref{eq:sp}) \citep[see][]{kud11}. 
Since we are interested in subcritical clouds, we take $\beta_0$ 
values less than $1$ in this paper.
Dimensional values can be obtained by specifying $\rho_0$ and $c_{s0}$.
In the case of  $c_{s0}=0.2$ km s$^{-1}$ and $n_0=\rho_0/m_n=10^4$ cm$^{-3}$
where $m_n=2.33 \times 1.67 \times 10^{-24} \rm{g}$, 
we obtain $H_0 \simeq 0.05$ pc, $t_0 \simeq 2.5 \times 10^5$ year, and 
$B_0 \simeq 40\,\mu$G if $\beta_0=0.25$.

The numerical methods are based on 
those of \citet{oga04} for solving the MHD equations 
and \citet{miy87b} for solving the self-gravity. 
The calculating area is taken to be $- 4\pi H_0 \leq x \leq 4\pi H_0$ and 
$0 \leq z \leq 4H_0$. 
The number of grid points is 
256 for the $x$-direction and 40 for the $z$-direction.
A mirror-symmetric boundary 
at $z=0$ and periodic boundaries in the $x$-directions are used. 
A mirror-symmetric boundary is also used at the upper boundary ($z=4H_0$)
for convenience except 
for the gravitational potential
(see details and discussion in \citet{kud11}). 
As it was done before \citep{kud07,kud08,kud11}, 
the ambipolar diffusion term is assumed to work only when the
density exceeds a certain value, $\rho_{cr}=0.3\rho_0$ to prevent
small time steps occurring outside of the low temperature cloud.

\section{Numerical Results and Semi-Analytical Formulation}

\subsection{Parameters} 

The summary of the models and parameters for the simulations are given in Table 1.
In this table, the values of the free parameters
$\beta_{0}$ and $v_{max}$ are listed.
The parameter $v_{max}$ is listed both normalized by the sound speed $c_{s0}$
and the initial Alfv\'en speed $v_{A0}$
on the midplane ($z=0$) of the clouds, 
where $v_{A0}=B_0/\sqrt{4\pi\rho_0}$.
The core formation time $t_{core}$,
which is characterized as the time when the maximum density attained 100 $\rho_0$,
is also listed. 
The definition of $t_{core}$ is the same as that in \citet{kud11}. 
Although $t_{core}$ is defined from a practical restriction of the numerical simulation, 
the center of the core always showed the features of runaway collapse at that time.
In the models A003 to A10, the amplitude of the initial velocity fluctuation $v_{max}$
is changed with the fixed $\beta_{0}=0.25$.
In the models B01 to B15, $v_{max}$ is changed with the fixed $\beta_{0}=0.09$.
In the models C1 to C7, $v_{max}$ is also changed with the fixed $\beta_{0}=0.49$.
In the models D3 to D15, $v_{max}$ is changed with a fixed $\beta_{0}=0.04$.
In addition to these models, 
the model E5 is also introduced, with the same parameters as those of model A5
except that the ambipolar diffusion is artificially switched off.

\subsection{General Properties} 

\begin{figure}
\epsscale{.80}
\plotone{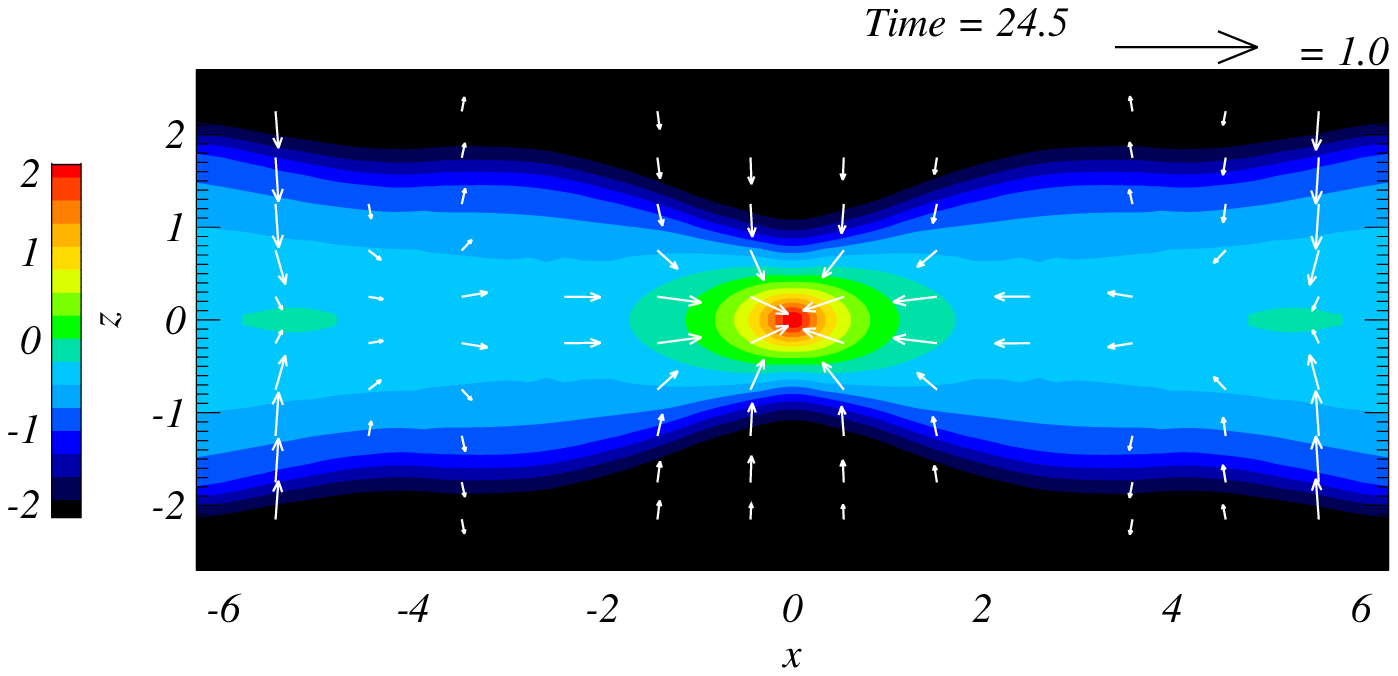}
\caption{Logarithmic density contours at $t/t_0=24.5$ for model A5.
Arrows show velocity vectors. \label{fig1}}

\epsscale{.80}
\plotone{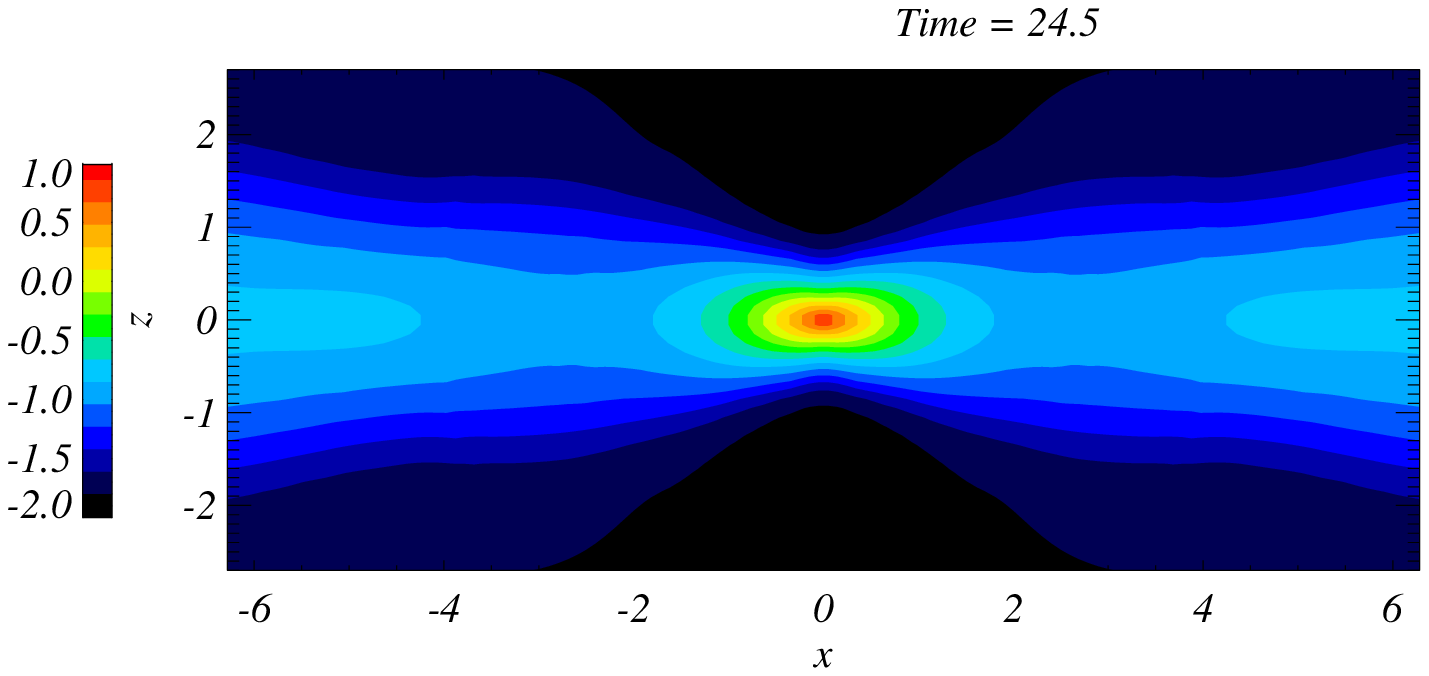}
\caption{Logarithmic plasma $\beta$ at $t/t_0=24.5$ for model A5.\label{fig2}}
\end{figure}

In Figure \ref{fig1}, the snapshot of the logarithmic density for the model A5
is shown as a color map, when the maximum density becomes greater than 
a hundred times of the initial density.
The vectors overlaid in Figure \ref{fig1} indicate the velocity.
Since the initial flow
converges the cloud toward $x=0$,
the figure shows that a collapsing core is located at the origin of the coordinate.
The shape of the core is approximately oblate whose apse line is 
along the $x$-direction in spite of the converging flow.
This is because the core formation does not occur during the first compression, 
but it occurs gradually in the oscillating cloud. We will see this process in Figure \ref{fig5} later.
The logarithmic plasma $\beta$ (the ratio of gas pressure to magnetic pressure)
for the model A5 is shown in Figure \ref{fig2} at the same time as that of Figure \ref{fig1}.
Figure \ref{fig2} shows that $\beta$ is greater than 1 near the center of the core, 
though the initial $\beta$ is $0.25$ on $z=0$ ($\beta_0=0.25$).
It means that the magnetic flux has been removed from the core 
and the mass-to-flux ratio of the collapsing core
is expected to be greater than 1 (see equation \ref{eq:bms}), i.e.,
the core is supercritical around the center.
We will see this feature again in the following discussion.

\begin{figure}
\epsscale{.80}
\plotone{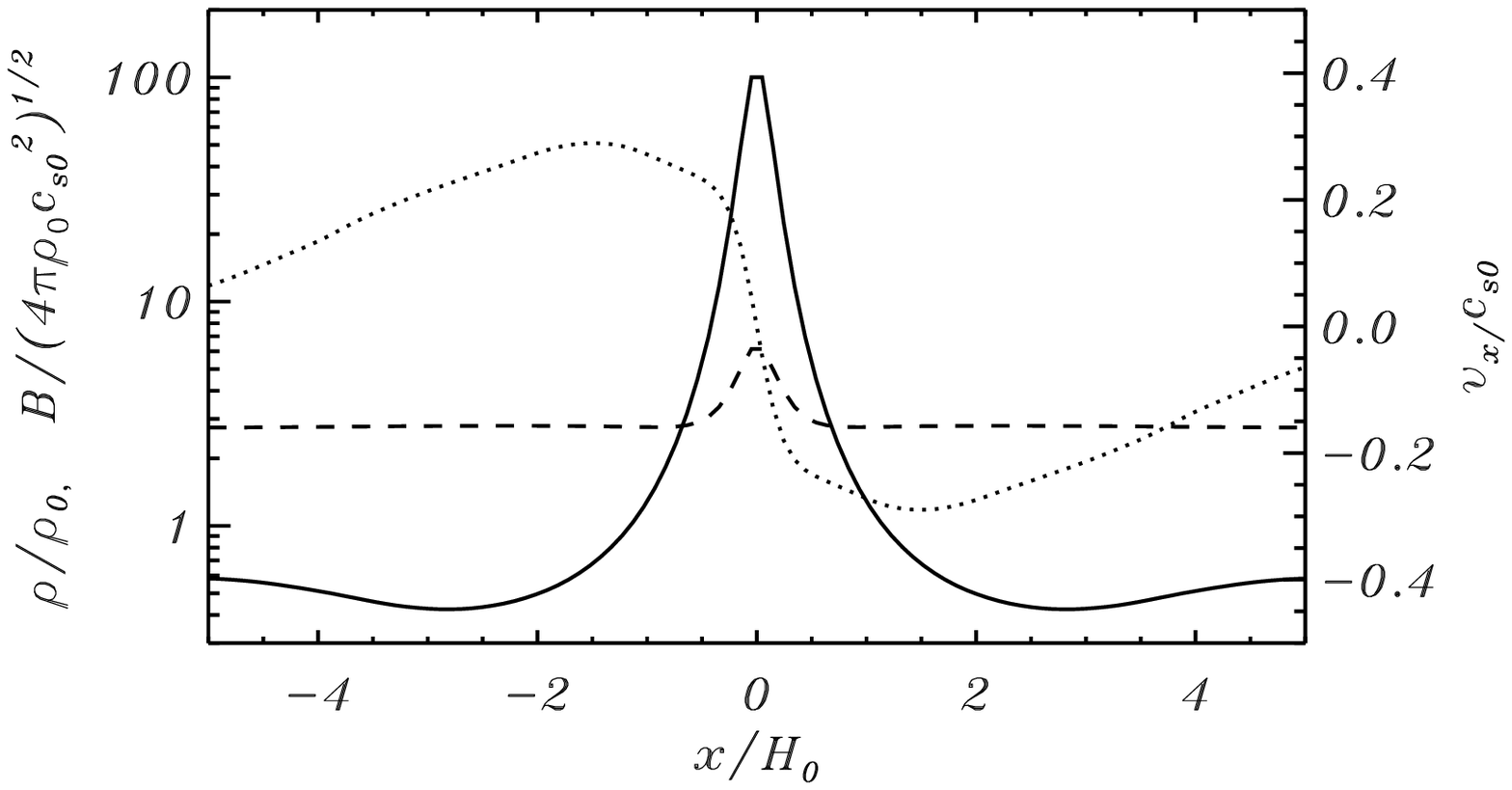}
\caption{ 
	Spatial profiles of density (the solid line), 
	strength of magnetic field (the dashed line), 
	and x-velocity (the dotted line) along $x$ on $z=0$ for model A5 at $t/t_0=24.5$.
\label{fig3}}

\epsscale{.80}
\plotone{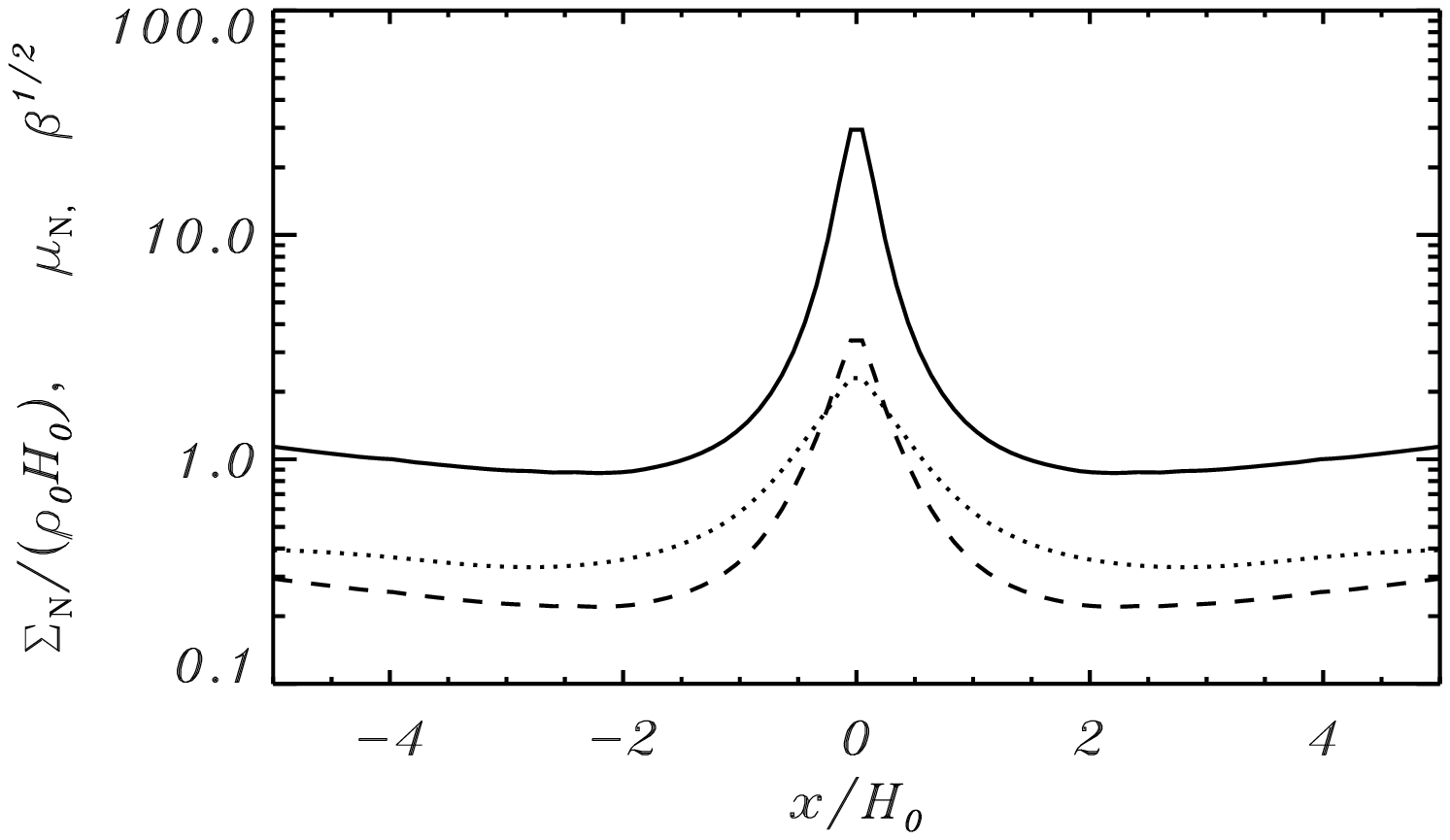}
\caption{ 
	Spatial profiles of surface density (the solid line), 
	and normalized mass-to-flux ratio ($\mu_N$: the dashed line) along $x$
	for model A5 at $t/t_0=24.5$.
The square root of plasma $\beta$ (the dotted line) along $x$ on $z=0$ is also plotted.
\label{fig4}} 
\end{figure}

\begin{figure}
\epsscale{.80}
\plotone{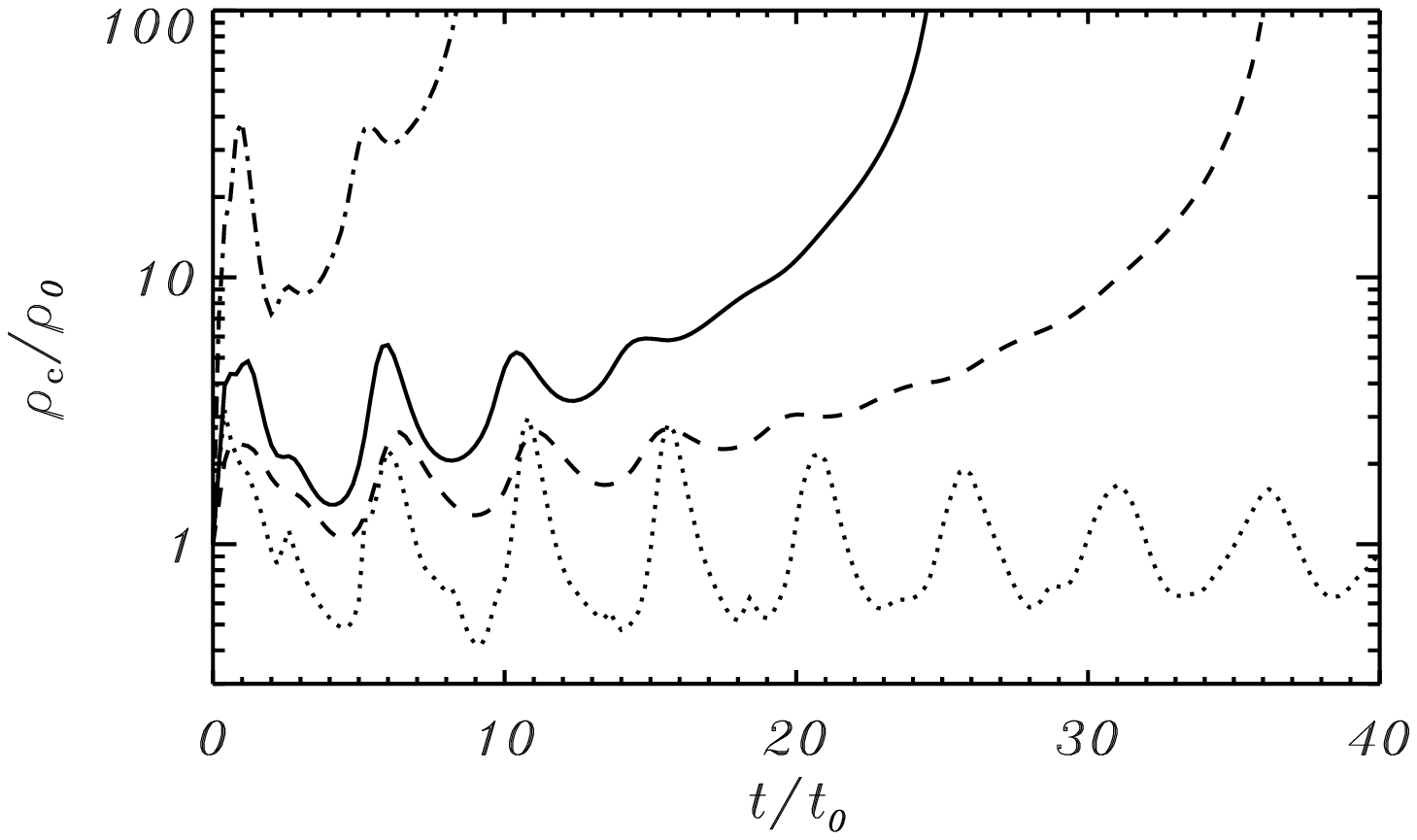}
\caption{ Time evolution of the density at $x=z=0$ ($\rho_c$). 
	The solid line shows the case of model A5.
	The dashed lines show the case of model A3, and
the dotted-dashed lines show the case of model A10.
	The dotted lines shows the case of model E5 in which the parameters are the
same as those of model A5 except that the ambipolar diffusion is switched off.
\label{fig5}}

\epsscale{.80}
\plotone{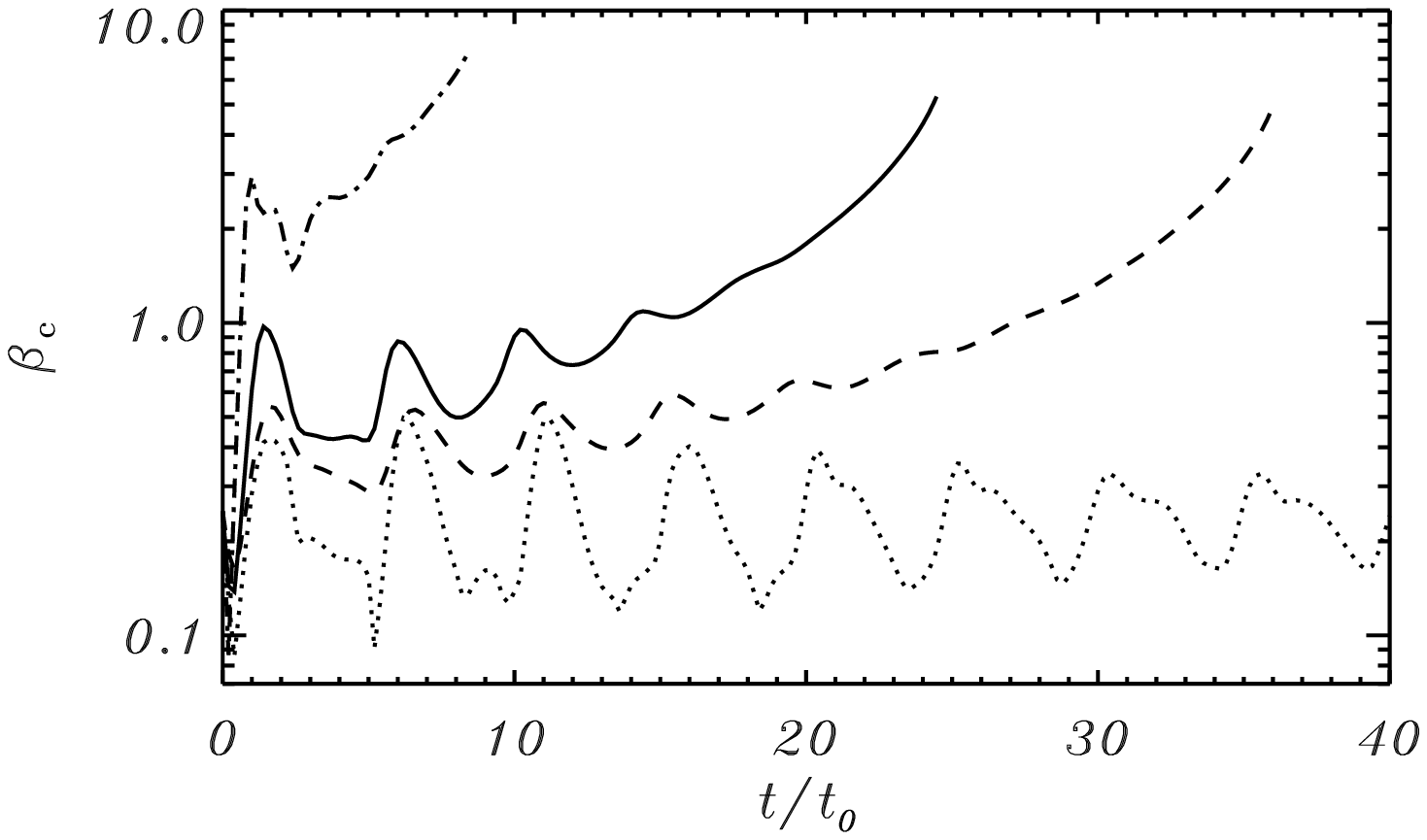}
\caption{ Time evolution of the plasma $\beta$ at $x=z=0$ ($\beta_c$).. 
	The solid line shows the case of model A5.
	The dashed lines show the case of model A3, and
the dotted-dashed lines show the case of model A10.
The dotted lines shows the case of model E5 in which the parameters are the
same as those of model A5 except that the ambipolar diffusion is switched off. 
\label{fig6}} 
\end{figure}

Figure \ref{fig3} shows the snapshot ($t=24.5t_0$) of the spatial profiles of density, 
magnetic field strength, and x-velocity along $x$ on $z=0$ for the model A5.
Both the density and magnetic field have peaks at $x=0$ where the flow
converges. 
The flow remains large scale and does not become turbulent in this
simulation.
Figure \ref{fig4} shows the snapshot ($t=24.5t_0$) of spatial profiles 
of surface density ($\Sigma_N$) and normalized mass-to-flux ratio ($\mu_N$) along $x$ 
for the same model A5. 
They are conventionally defined 
in the following equations:
\begin{equation}
	\Sigma_N \equiv \int^{z_{B}}_{-z_{B}} \rho dz,
\end{equation}

\begin{equation}
	\mu_N \equiv 2\pi G^{1/2} \frac{\Sigma_N}{|B_z(z=0)|},
\end{equation}
where $z_{B}=4H_0$ is the upper boundary value in the calculating area, 
and $B_z(z=0)$ is the $z$ component of the magnetic field at $z=0$.
The normalized mass-to-flux ratio is greater than 1 near the peak 
of the surface density. This means that the region has 
evolved from a subcritical to supercritical state
through the removal of magnetic flux.
In Figure \ref{fig4}, the square root of plasma $\beta$ at $z=0$ is also plotted. 
It shows that $\beta^{1/2}$ follows a similar profile as that of the normalized 
mass-to-flux ratio, and $\beta$ is nearly greater than 1 where $\mu_N$ is 
greater than 1.
Since $\beta$ is locally defined in the simulation, it is convenient to use $\beta$
as an approximate indicator of the magnetic criticality of the cloud, 
although it is not exactly the same as $\mu_N$.

Figure \ref{fig5} shows the time evolution of the density at the location of 
the core center ($\rho_c$), i.e., $(x,z)=(0,0)$.
The solid line shows the case of model A5.
The density increases during the first compression of the cloud. 
Next, it bounces and undergoes oscillations. 
Finally, the dense region goes into runaway collapse after $t/t_0 \sim 20$.
Figure \ref{fig6} shows the time evolution of $\beta$ at the the core center ($\beta_c$). 
In the very initial stage of the compression ($t/t_0 < 1$), $\beta_c$ rapidly decreases, 
but it increases soon as the density reaches a peak around $t/t_0 \sim 1$. 
After that, it decreases and shows oscillations as seen in the time evolution of the density.
Eventually, it gradually increases with oscillations and 
becomes greater than $1$ when the runaway collapse happens.
Since $\beta_c$ approximately equals the square of the mass-to-flux ratio,
the runaway collapsing core is expected to be supercritical.
Figure \ref{fig6}, as well as Figure \ref{fig2} and Figure \ref{fig4}, 
shows that $\beta_c$ can be 
a good indicator of the mass-to-flux ratio during the oscillations, even though the 
gravitational equilibrium of the gas is not exactly established at the final 
stage of the simulation.
The first rapid decrease of $\beta_c$ ($t/t_0 < 1$) could be a 
result of the non-equilibrium of the gas along the magnetic field.
The dotted lines in Figures \ref{fig5} and \ref{fig6} show the case of model E5
in which the parameters are the same as those of model A5 except 
that the ambipolar diffusion is artificially switched off.
When the ambipolar diffusion is switched off, the density and $\beta_c$ do not show
the feature of collapse while they do show oscillations. This also indicates that the
collapse is caused by the removal of the magnetic flux through the ambipolar diffusion.
In Figures \ref{fig5} and \ref{fig6}, the dashed lines show the case of model A3 
whose initial converging flow speed is smaller than that of model A5,
and
dotted-dashed lines show the case of the model A10
whose initial flow speed is larger than that of the model A5.
The overall features of the time evolutions are qualitatively similar,
though the core formation time is shorter when the initial flow speed is greater.
When the initial flow speed is high enough (for the model A10), $\beta_c$ becomes greater
than $1$ at the first peak. In that case, it rebounds a little bit, 
but goes into runaway collapse very soon. 

\subsection{Core Formation Time}

Figure \ref{fig7} shows the core formation time as a function of the initial flow speed $v_{max}$. 
The flow speed is normalized by the sound speed.
The filled circles show the case of $\beta_0=0.25$ (the models A003 to A10).
The open squares show the case of $\beta_0=0.09$ (the models B01 to B15).
The open triangles show the case of $\beta_0=0.49$ (the models C1 to C7).
The filled triangles show the case of $\beta_0=0.04$ (the models D3 to D15).
As the flow speed increases, the core formation time decreases for each 
set of $\beta_0$ values.
The case of the smaller $\beta_0$, i.e., the smaller initial mass-to-flux ratio,
shows a longer core formation time for the same initial flow speed.
The flow speed normalized by the sound speed may be useful when the results
are compared with observations.
On the other hand, in figure \ref{fig8}, we plot the core formation time as a function of $v_{max}$, 
like in figure \ref{fig7}, but normalized by the initial Alfv\'en speed on the midplane of the clouds
($v_{A0}$).
In figure \ref{fig8}, all results fall on the same line regardless of the 
value of $\beta_0$. The core formation time is approximately the same for
the same initial flow speed if it is normalized by the  Alfv\'en speed.
The dashed line shows the semi-analytic result which is obtained in the
next subsection.
The core formation time is in inverse proportion to the square of 
$v_{max}$, when $v_{max}$ is greater than $v_{A0}$.
We will have a detailed discussion of this result in the next subsection.

The core formation time is also shown in figure \ref{fig9}
as a function of 
$\rho_{peak}$, defined as the value of 
the density peak during the first compression
in the time evolution of the central density at $x=0$ and $z=0$.
As we have discussed in our previous paper (Figure $16$ in \citet{kud11}), 
the core formation time is proportional to $\rho_{peak}^{-0.5}$ 
when $\rho_{peak}/\rho_0$ is enough greater than $1$. 
The dashed line shows the semi-analytic result obtained in the next subsection,
which shows $t_{core} \propto \rho_{peak}^{-0.5}$. 
Figure \ref{fig9} shows the relation clearly, and the figure also indicates that the relation
does not depend on $\beta_0$, i.e., the initial mass-to-flux ratio.
The result is also discussed in the next subsection in detail.

\subsection{Semi-Analytical Formulation} 

In this subsection, we find the semi-analytical formulas
of core formation time and compare them with the numerical results
in the previous subsections.
Since figure \ref{fig5} and figure \ref{fig6} shows the oscillation of the cloud, 
we considered the cloud to be in an approximate force balance 
for a time average of the period.
When we assume force balance between the magnetic force and gravity along the $x$-axis 
in the cylindrical subcritical cloud \citep{mou99},
\begin{equation}
\rho \frac{GM_l}{L} \sim \frac{B^2}{8\pi} \frac{1}{L},
\end{equation}
where
\begin{equation}
M_l \sim \pi H L \rho 
\end{equation}
is the line mass along the cylinder, $L$ is the half-size of the cloud in the $x$-direction, 
and $H$ is the half thickness of the cloud in $z$-direction.
The magnetic field is derived from these two equations as 
\begin{equation}
	B^2 \sim 8\pi\rho G M_l \sim 8\pi^2 G \rho^2 H L.
\end{equation}
From the equation (\ref{eq:ind}), 
the ambipolar diffusion time ($\tau_{AD}$) is estimated to be
\begin{equation}
	\tau_{AD} \sim \frac{4\pi}{\gamma} \frac{L^2 \rho^{3/2}}{B^2} 
		\sim \frac{1}{2\pi\gamma G} \frac{L}{H} \rho^{-1/2}.	
\end{equation}
Since the most unstable wave length of the subcritical cloud is about $4\pi H$, 
we take
\begin{equation}
	L \sim 4\pi H \times \frac{1}{2}.
\end{equation}
Therefore, we get
\begin{equation}
	\tau_{AD} \sim \frac{1}{\gamma G}\rho^{-1/2}.
	\label{eq:ambdt}
\end{equation}
This is the same equation that was derived in \citet{mou99}. 
Although we derive it for cylindrical clouds, the result does not depend on the
geometry of the cloud.
The dashed line in figure \ref{fig9} is drawn from the equation (\ref{eq:ambdt}),
when we use $\rho \sim \rho_{peak}$ approximately.
The figure \ref{fig9} shows that the scaling relation agrees with the result of the numerical simulations.
The relation also indicates that it does not depend on the initial mass-to-flux ratio.

Next, we consider the local pressure balance during the compression by the large scale flows.
Since the cloud is subcritical, the thermal pressure can be negligible. Then, the pressure balance
along $x$ axis
will be
\begin{equation}
	H \frac{B^2}{8\pi} \sim H_0 \left(\rho_0 {v_{t0}}^2 + \frac{{B_0}^2}{8\pi} \right),
	\label{eq:pb}
\end{equation}
where $v_{t}$ is the nonlinear flow speed in the cloud and the subscripts $0$ mean
the values before the compression.
When we assume that there is sufficient time for the vertical density structure 
to be put back into hydrostatic equilibrium, i.e., $t/t_0 > 1$, $H$ is estimated to be
\begin{equation}
	H \sim \frac{c_s}{\sqrt{2\pi G \rho}}. 
\end{equation}
Therefore, the equation (\ref{eq:pb}) becomes 
\begin{equation}
	\frac{1}{\sqrt{\rho}} \frac{B^2}{8\pi} \sim \frac{1}{\sqrt{\rho_0}} \left(\rho_0 {v_{t0}}^2 + \frac{{B_0}^2}{8\pi} \right),
	\label{eq:pb2}
\end{equation}
in the isothermal gas, and this leads to
\begin{equation}
	\left( \frac{\rho}{\rho_0} \right)^{1/2} \sim 
	\left(v_{t0}^2 + \frac{{B_0}^2}{8\pi\rho_0} \right) \left(\frac{B^2}{8\pi\rho} \right)^{-1}.
	\label{eq:pb3}
\end{equation}
When the ambipolar diffusion time is longer than the compression time, 
flux-freezing can be a good approximation during the compression, i.e.,
\begin{equation}
	\frac{B}{\Sigma} \sim \frac{B_{0}}{\Sigma_{0}},
	\label{eq:flfz}
\end{equation}
where $\Sigma$ is the surface density. Since $\Sigma \sim 2\rho H$, 
equation (\ref{eq:flfz}) is compatible with
\begin{equation}
	\frac{B}{\rho^{1/2}} \sim \frac{B_0}{{\rho_0}^{1/2}}
	\label{eq:fluxf}
\end{equation}
in our approximation. Finally, equation (\ref{eq:pb3}) then becomes
\begin{equation}
	\left( \frac{\rho}{\rho_0} \right)^{1/2} \sim
	2 \left( \frac{v_{t0}}{v_{A0}} \right)^2 +1,
	\label{eq:d-enhance}
\end{equation}
where 
\begin{equation}
          v_{A0}^2 = \frac{{B_0}^2}{4\pi\rho_0} 
\end{equation}
is the square of the Alfv\'en speed of the cloud.
By using this equation, the equation for the ambipolar diffusion time (\ref{eq:ambdt}) 
is expressed using the flow speed as
\begin{equation}
	\tau_{AD} \sim \frac{1}{\gamma G}{\rho_0}^{-1/2} \left[2 \left( \frac{v_{t0}}{v_{A0}} \right)^2 +1 \right]^{-1}.
	\label{eq:ambdtv}
\end{equation}
The dashed line in figure \ref{fig6} is drawn from the equation (\ref{eq:ambdtv}),
when we use $\displaystyle v_{t0} \sim 0.5 v_{max}$ as an average speed.
The figure \ref{fig6} shows that the scaling relation agrees with the results of the numerical simulations.
When the  flow speed is comparable to the  Alfv\'en speed, the ambipolar diffusion time
is estimated to be
\begin{equation}
	\tau_{AD} \sim \frac{1}{3} \frac{1}{\gamma G}{\rho_0}^{-1/2},
\end{equation}
which means the the ambipolar diffusion time is about three times shorter than 
that without the disturbance.
When the flow speed is greater than the Alfv\'en speed, 
\begin{equation}
	\tau_{AD} \propto v_{t0}^{-2},
\end{equation}
which means that the ambipolar diffusion time
is inversely proportional to the square of the nonlinear flow speed in 
a subcritical magnetic cloud.

\section{Summary and Discussion}

\begin{figure}
\epsscale{.80}
\plotone{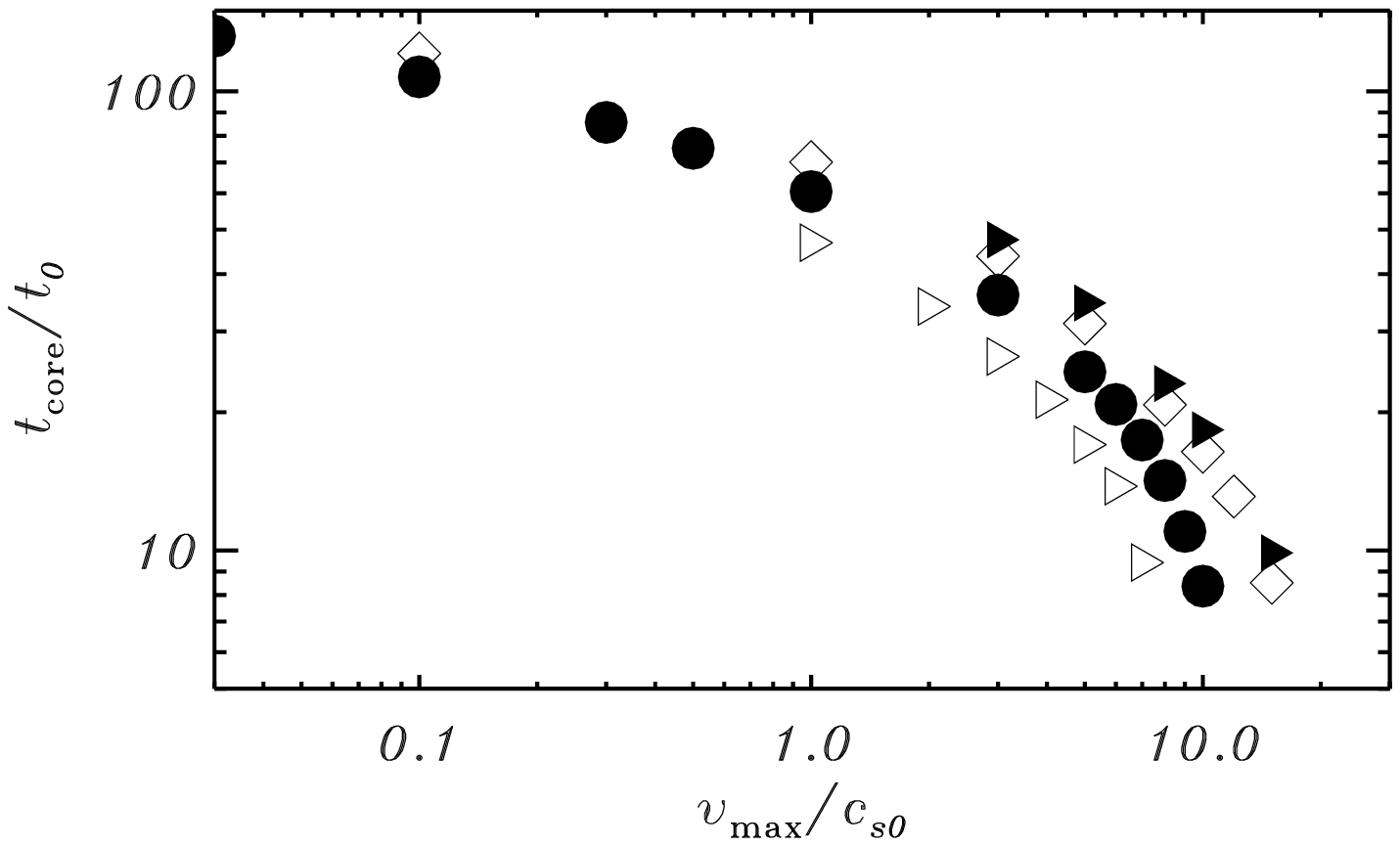}
\caption{ Core formation time as a function of initial
	amplitude of flow speed normalized by the sound speed
	in the clouds.
	The filled circles show the case of $\beta_0=0.25$.
The open squares show the case of $\beta_0=0.09$.
The open triangles show the case of $\beta_0=0.49$.
The filled triangles show the case of $\beta_0=0.04$.
\label{fig7}}

\epsscale{.80}
\plotone{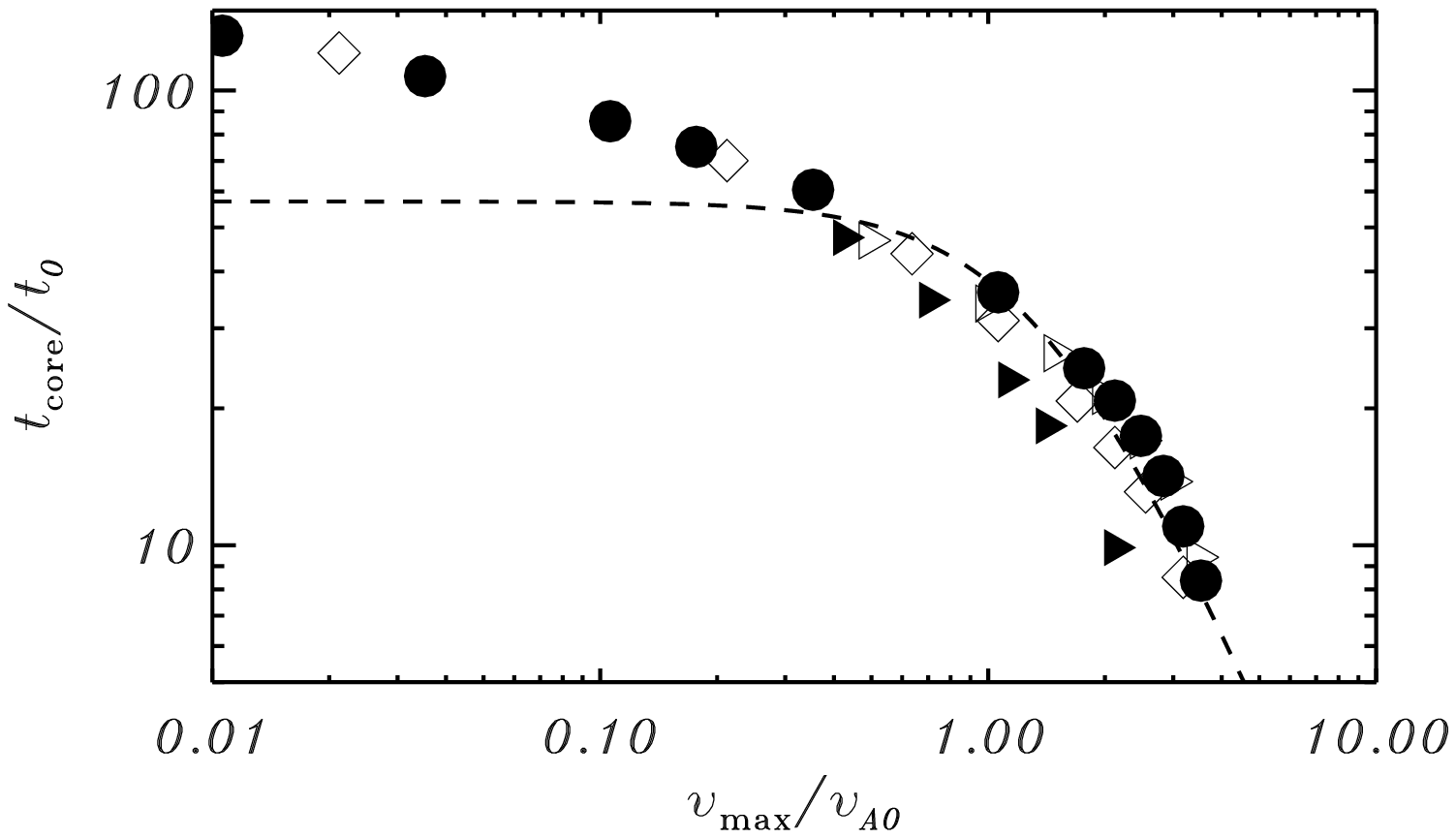}
\caption{ Core formation time as a function of initial
	amplitude of flow speed normalized by initial Alfv\'en 
	speeds on the midplanes in clouds.
	The filled circles show the case of $\beta_0=0.25$.
The open squares show the case of $\beta_0=0.09$.
The open triangles show the case of $\beta_0=0.49$.
The filled triangles show the case of $\beta_0=0.04$.
The dashed line is drawn from equation (\ref{eq:ambdtv}),
        using $v_{t0} \sim 0.5 v_{max}$ as an average speed.
\label{fig8}}
\end{figure}

\begin{figure}
\epsscale{.80}
\plotone{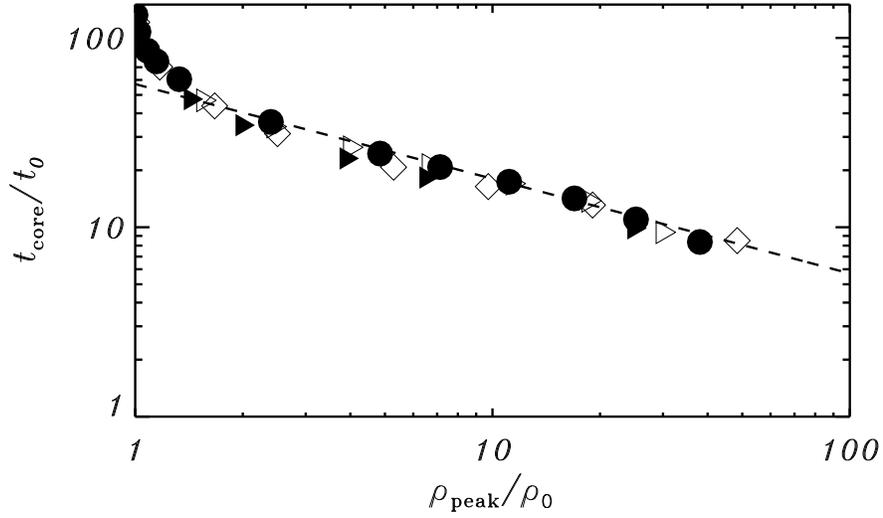}
\caption{ Core formation time as a function of the density peak ($\rho_{peak}$)
	during the first compression in its time evolution.
	The filled circles show the case of $\beta_0=0.25$.
The open squares show the case of $\beta_0=0.09$.
The open triangles show the case of $\beta_0=0.49$.
The filled triangles show the case of $\beta_0=0.04$.
The dashed line is drawn from equation (\ref{eq:ambdt}),
	using $\rho \sim \rho_{peak}$. 
\label{fig9}}
\end{figure}

Since the core formation time ($t_{core}$) is almost comparable to 
the ambipolar diffusion time in a subcritical magnetic cloud,
the relations
\begin{equation}
	t_{core} \sim \tau_{AD} \sim \frac{1}{\gamma G}\rho^{-1/2},
	\label{eq:cftd}
\end{equation}
and
\begin{equation}
	t_{core} \sim \tau_{AD} \sim \frac{1}{\gamma G}{\rho_0}^{-1/2} \left[2 
	\left( \frac{v_{t0}}{v_{A0}} \right)^2 +1 \right]^{-1} 
	\label{eq:cftv}
\end{equation}
are consistent with the results of the numerical simulations in figure \ref{fig8} and figure \ref{fig9}
as long as the turbulent velocities ($v_{t0}$) are large enough to compress the cloud,
using $\rho \sim \rho_{peak}$ as a representative density 
and $v_{t0} \sim 0.5 v_{max}$ as an average speed.
When the observed turbulent speed is comparable to the Alfv\'en speed,
the core formation time is reduced by about $3$ times from the
usual ambipolar diffusion time.
Furthermore, the core formation time is considerably reduced 
when the turbulent speed is larger than the Alfv\'en speed, i.e., the core formation time
is inversely proportional to the square of the turbulent speed. 

The equations (\ref{eq:ambdt}) and (\ref{eq:cftd}) are the same as those
considered by \citet{mou99}. Although \citet{mou99} considered the static force balance
of a subcritical cloud, we showed here that similar relations are applicable to
the dynamically oscillating cloud. In addition to the global force balance, 
a local pressure balance between the magnetic and dynamic 
pressure is assumed in the dynamically compressible subcritical cloud 
to obtain the velocity dependence of the ambipolar diffusion time, i.e.,
the equations (\ref{eq:ambdtv}) and (\ref{eq:cftv}). 
When equation (\ref{eq:ambdtv}) is derived, the relation $B \propto \rho^{1/2}$
is used in the turbulent flows assuming magnetic flux freezing 
as in equation (\ref{eq:flfz}) and  (\ref{eq:fluxf}) .
\citet{fat02} studied ambipolar diffusion with the limit of long-wave fluctuations
and suggested that there is no net enhancement of the ambipolar diffusion 
when the density fluctuation depends on the magnetic field fluctuation 
as $B \propto \rho^{1/2}$.
In our case, however, the diffusion is enhanced even when the relation
is satisfied. 
The difference possibly may come from the fact that \citet{fat02}
used a one-dimensional slab model in which the self-gravitational force does not
change during the compression along the magnetic field lines if
the surface density is constant.
In our two-dimensional slab model, the compression perpendicular 
to the magnetic field enhances the self-gravitational force 
as well as the density.
Our result essentially demonstrates the same result as that of the \lq\lq gravitationally driven 
ambipolar diffusion" discussed in \citet{mou99} etc, but the density enhancement along with 
that of the self-gravity is induced by the large-scale compressive turbulent flows.

We showed that the large-scale turbulence associated with density 
enhancement is an important factor for the fast core formation 
in turbulent subcritical clouds.
In our analysis, the turbulence whose scale is larger than about $4\pi H_0$,
which is about $0.6$ pc for typical molecular clouds,
is needed to get an effective fast ambipolar diffusion.
Although the origin of the turbulence in molecular clouds has not yet been identified,
the scale of the energy source of the turbulence is supposed to be larger 
than $4\pi H_0 \sim 0.6$ pc to realize the fast core formation considered in this paper.
By comparison,
the small-scale turbulence is also considered to be an important factor 
for the turbulent diffusion processes.
For example, the fast ambipolar diffusion rate in a turbulent medium was studied 
by \citet{zwe02} to explain low magnetic field
strength in dense interstellar gas.
\citet{lea13} pointed out that the \lq\lq turbulent reconnection diffusion"
is also effective to remove the magnetic flux from subcritical cores.
These processes would be more efficient if the energy source of the turbulence 
in molecular clouds is originated in smaller scales.
Since the large-scale turbulent flows even in two- or 
three-dimensional spaces tend to compress the cloud in a one-dimensional manner 
to make filamentary structures, as seen in \citet{bas09} and \citet{kud11},
our results can explain generally how core formation is accelerated by
the nonlinear large-scale turbulent flows in subcritical magnetic clouds.
The results obtained in this paper would 
be applicable to
a filamentary molecular cloud \citep{and10} if it is created 
by a large-scale compression \citep{per12}.
Once the filamentary structure becomes supercritical by the accelerated
ambipolar diffusion, the fragmentation along the filament would occur 
on a dynamical timescale to produce several cores.

When $v_{max}$ is much smaller than the the Alfv\'en speed, 
figure \ref{fig8} shows that the equation (\ref{eq:ambdtv}) does 
not fit to the numerical results well.
This comes from the fact that the equation (\ref{eq:ambdt}) also does not fit 
to the numerical results well in figure \ref{fig9}, when the density peak is small.
When the density peak is very small, the assumption of the force balance between
gravity and magnetic force for the density peak, which is used to
derive equation  (\ref{eq:ambdt}), may not be acceptable. 
Thus, the formula may not be well applied to the quasi-linear regime of the compression, 
although it is applicable to the nonlinear regime in which we are mostly interested 
in this paper.
When $v_{max}$ is much greater than the Alfv\'en speed,
the collapse may happen more quickly during the compression.
If the flow speed is large enough that vertical hydrostatic equilibrium
along the magnetic field cannot be attained, then 
$\rho \propto \Sigma \propto L^{-1}$ and 
$\tau_{AD} \propto L^{5/2} \propto \rho^{-5/2}$ \citep{elm07,kud08}.
(The scaling law of $\tau_{AD}$ is derived from 
equation (\ref{eq:tauadb}) by using $H \sim H_0$.)
In this case, the ambipolar diffusion can occur more quickly 
when the compression leads to large values of $\rho$.
This process would be happening 
for $t/t_0 < 1$ even in our results, although the overall timescale 
of the core formation is dominated by the time after the vertical hydrostatic equilibrium is reestablished.
The simulation with greater $v_{max}$ was not successful in our case
because the density after the first compression becomes so great that
the simulation does not have enough spatial resolution for 
the high density self-gravitating cores.
Adaptive mesh refinement or nested grid techniques might be needed 
for the further investigation in the case of greater $v_{max}$.
However, the turbulent flows in molecular clouds
are observed to not be highly super Alfv\'enic
(see the fit to data in \citet{bas00}). 
The parameters we used in this paper can be considered to be within a proper 
range for typical molecular clouds.

In the case of $\beta_0=0.04$, in figure \ref{fig8},
the core formation time seems to be slightly smaller than 
what is expected from the semi-analytic formula. 
This may be caused by the fact 
that the approximation of flux-freezing during the compression, 
which is used in equation (\ref{eq:flfz}),
is not a good approximation 
when the parameter $\beta_0$ is small.
The local ambipolar diffusion time ($\tau_{AD}$) is estimated to be
\begin{eqnarray}
\tau_{AD} & \sim & \frac{4\pi}{\gamma} \frac{L^2 \rho^{3/2}}{B^2} \\
          & \sim & \frac{4\pi}{\gamma} \frac{H_0^2 \rho_0^{3/2}}{B_0^2} 
	                   \left(\frac{L}{H}\right)^2 \left(\frac{\rho}{\rho_0}\right)^{-1/2} \\
          & \sim & \frac{1}{2\gamma\sqrt{2\pi G}} \beta_0 
\left(\frac{L}{H}\right)^2 \left(\frac{\rho}{\rho_0}\right)^{-1/2}t_0,
\label{eq:tauadb}
\end{eqnarray}
when we use equation (\ref{eq:flfz}).
When $\beta_0 \simeq 0.04$, $L \sim 2\pi H$ and $\rho \sim 10 \rho_0$, 
we can get $\tau_{AD} \sim 2t_0$. 
This means that the local ambipolar diffusion time
after the compression becomes nearly comparable to the compression time when $\beta_0=0.04$.
If the local ambipolar diffusion time is comparable to the compression time,
the density can be much more enhanced than estimated 
from equation (\ref{eq:d-enhance}) for the same velocity 
because the magnetic pressure after the compression is reduced by the diffusion. 
This density enhancement leads to a smaller core formation time through 
equation (\ref{eq:ambdt})
that seems to be satisfied even in the case of $\beta_0=0.04$ from Figure \ref{fig9}.
The parameter $\beta_0=0.04$ corresponds to 
$\mu_S = 0.2$.
The case of even lesser initial mass-to-flux ratio may result 
in a shorter core formation time 
than is estimated from equation (\ref{eq:ambdt}).
This should be confirmed in the future, and will require greater
spatial resolution because a stronger magnetic field generally
requires greater density near the core center
before the onset of runaway collapse.

From Table 1 and Figure \ref{fig7}, the dimensional core formation time 
is estimated to be 
about $2 \times 10^6$ year, when the flow speed in the cloud is 10 times 
greater than the sound speed, in the case that the initial mass-to-flux ratio 
is about half of the critical value (i.e., $\beta_0=0.25$). 
This core formation time is consistent with those of the previous three-dimensional 
simulations with turbulent flows \citep{kud11}, although 
the core formation time of the present simulation is slightly longer than 
that of the three-dimensional simulation with the same initial flow speed.
This is because that the flow speeds in the three dimensional 
simulations were the averages of the turbulent flows, and the flow components with
speeds greater than the average could slightly shorten the core formation times.
In both cases, our simulations show that the core formation time 
is estimated to be the order of a few to several $\times 10^6$ years
for the subcritical molecular clouds whose initial mass-to-flux ratios
are about half of the critical value given that initial large-scale flow
speeds (whether they are turbulent or not) are $3-10$ times greater than the sound speed.

\acknowledgments
Numerical simulations were performed on the SX-9 
and on the PC cluster at 
the Center for Computational Astrophysics in 
National Astronomical Observatory of Japan. 
S.B. is supported by a Discovery Grant from the Natural Sciences and 
Engineering Research Council (NSERC) of Canada.

\clearpage

\clearpage

\begin{deluxetable}{cccccc}
\tabletypesize{\scriptsize}
\tablecaption{Model and Parameters}
\tablewidth{0pt}
\tablehead{
\colhead{Model} & \colhead{$\beta_{0}$} &  \colhead{$v_{max}/c_{s0}$} & \colhead{$v_{max}/v_{A0}$} &
	\colhead{$t_{core}/t_0$} & notes 
}
\startdata
A003 & 0.25 & 0.03 & 0.011 &   132  &  \\
A01 & 0.25 & 0.1  & 0.035 &   108  & \\
A03 & 0.25 & 0.3  & 0.11 &   85.6  & \\
A05 & 0.25 & 0.5  & 0.18 &   75.2  & \\
A1 & 0.25 & 1.0  & 0.35 &   60.5  & \\
A3 & 0.25 & 3.0  & 1.1 &   36.0  & \\
A5 & 0.25 & 5.0  & 1.8 &   24.5 &  \\
A6 & 0.25 & 6.0  & 2.1 &   20.8 &  \\
A7 & 0.25 & 7.0  & 2.5 &   17.4 &  \\
A8 & 0.25 & 8.0  & 2.8 &   14.2 &  \\
A9 & 0.25 & 9.0  & 3.2 &   11.0 &  \\
A10 & 0.25 & 10.0  & 3.5 &  8.36 & \medskip  \\
B01 & 0.09 & 0.1  & 0.021 &   121 &   \\
B1 & 0.09 & 1.0  & 0.21 &   70.1 &  \\
B3 & 0.09 & 3.0  & 0.64 &   43.8 &  \\
B5 & 0.09 & 5.0  & 1.1 &   31.2 &  \\
B8 & 0.09 & 8.0  & 1.7 &   20.8 &  \\
B10 & 0.09 & 10.0 & 2.1 &   16.4 &  \\
B12 & 0.09 & 12.0 & 2.5 &   13.1 &  \\
B15 & 0.09 & 15.0 & 3.2 &   8.50 &  \medskip \\
C1 & 0.49 & 1.0  & 0.50 &   46.8 &  \\
C2 & 0.49 & 2.0  & 1.0 &   34.0 &  \\
C3 & 0.49 & 3.0  & 1.5 &   26.5 &   \\
C4 & 0.49 & 4.0  & 2.0 &   21.3 &  \\
C5 & 0.49 & 5.0  & 2.5 &   17.0 &  \\
C6 & 0.49 & 6.0  & 3.0 &   13.8 &  \\
C7 & 0.49 & 7.0  & 3.5 &   9.41 &  \medskip \\
D3 & 0.04 & 3.0  & 0.42 &   47.5 &  \\
D5 & 0.04 & 5.0  & 0.71 &   34.6 &  \\
D8 & 0.04 & 8.0  & 1.1 &   23.1  &  \\
D10 & 0.04 & 10.0  & 1.4 &   18.3 &  \\
D15 & 0.04 & 15.0  & 2.1 &   9.88 &  \medskip \\
E5 & 0.25 & 5.0  & 1.8 &   $-$ & no ambipolar diffusion \\
\enddata
\tablecomments{
$\beta_{0}$ is the initial plasma $\beta$ at $z=0$,
which represents the square of initial mass-to-flux ratio
(see equation (\ref{eq:bms})). $v_{max}$ is the amplitude of 
the initial velocity fluctuation. $t_{core}$ is the time 
of collapsing core formation. 
$c_{s0}$ and $v_{A0}$ are initial sound and Alfv\'en speeds,
respectively.
}
\end{deluxetable}

\end{document}